\documentclass[twocolumn,english,prl,aps,superscriptaddress]{revtex4-2}
\usepackage[T1]{fontenc}%\usepackage[utf8]{inputenc}
\usepackage{color}
\usepackage{babel}
\usepackage{mathtools}
\usepackage{bm}
\usepackage{graphicx}
\usepackage{esint}
\usepackage{amsfonts}
\usepackage{graphicx}
\usepackage{float}
\usepackage{subfigure}
\usepackage{amsthm}
\usepackage{amssymb}
\usepackage{physics}
\usepackage[ruled,vlined,linesnumbered]{algorithm2e}
\usepackage[pdftex,colorlinks=true, linkcolor = blue, citecolor=blue, urlcolor=blue, 
bookmarksnumbered=true, bookmarksopen=true]{hyperref}
\usepackage{breakurl}
\usepackage{orcidlink}

\begin{document}

\title{Evaluation of the quantum Fisher information for large systems via an ancilla}

\author{Qi Liu}
\affiliation{School of Physics, Huazhong University of Science and Technology, 
Wuhan 430074, China}

\author{Xin-Yu Hao}
\affiliation{Center for Theoretical Physics and School of Physics and Optoelectronic 
Engineering, Hainan University, Haikou 570228, China}

\author{Jing Liu\orcidlink{0000-0001-9944-4493}}
\email{jing.liu@hainanu.edu.cn}
\affiliation{Center for Theoretical Physics and School of Physics and Optoelectronic 
Engineering, Hainan University, Haikou 570228, China}
\affiliation{School of Physics, Huazhong University of Science and Technology, 
Wuhan 430074, China}

\begin{abstract}
Quantum Fisher information (QFI) is a fundamental quantity in quantum information and quantum mechanics, and evaluating it in various 
physical systems has become an important research topic. However, its evaluation in large systems still faces many challenges. Hereby we 
propose an ancilla-based protocol for evaluating the QFI of large systems and establish the conditions under which this protocol can 
``perfectly'' evaluate the QFI in theory. As a demonstration, the protocol is applied to quantum circuits to evaluate the QFI of the 
Greenberger-Horne-Zeilinger state.
\end{abstract}

\maketitle

After the development over 60 years, quantum Cram\'{e}r-Rao bound (or Helstrom bound)\,\cite{Helstrom1976,Holevo1982,Toth2014,Liu2020} has 
become the most important theoretical tool in quantum parameter estimation and quantum metrology. In this bound, the quantum Fisher information 
(QFI) is the core quantity to depict the ultimate theoretical precision limit. For a density matrix $\rho(x)$ with $x$ a parameter, the corresponding QFI 
can be calculated by $F(\rho)=\mathrm{Tr}(\rho L^2)$. Here $F(\rho)$ is the QFI of $\rho$ and $L$ is the symmetric logarithmic derivative (SLD) 
operator determined by the equation $\partial_x \rho=(\rho L+L\rho)/2$. The QFI can be saturated by the classical Fisher information (CFI) with a 
theoretical optimal measurement. For a set of discrete probability distribution $\{p_i\}$, the corresponding CFI reads 
$I(\{p_i\})=\sum_i (\partial_x p_i)^2/p_i$.  

Nowadays, the QFI is important not only as a tool for assessing the ultimate precision limit, but also because of its connection to many important features 
and processes in quantum physics, such as entanglement\,\cite{Toth2014} and phase transitions\,\cite{Mazza2026}, which elevates it to a physical quantity 
independent of the precision-measurement scenario and makes it worth studying in its own right. However, the challenge is that the QFI is not always 
experimentally accessible. Hence, evaluating the QFI via various approaches becomes crucial in both theory and practice. 

Currently, several protocols have been proposed and discussed for the evaluation of the QFI in different scenarios\,\cite{Ozawa2018,Yu2019,Tan2019,
Ding2022,Macri2016,Kolodrubetz2013,Zhang2023,Yu2021,Rath2021,Zhang2025,Zhang2026,Yu2022,Cerezo2021}. One classical way to evaluate the 
QFI is to measure some lower bounds of the QFI which is experimentally accessible, including the sub-QFI given by Cerezo et al.\,\cite{Cerezo2021}, the 
converging lower bounds given by Rath et al.\,\cite{Rath2021}, and the Krylov lower bounds given by Zhang and Tong\,\cite{Zhang2025}. In the scenario 
of large systems, the direct measurement of these lower bounds would be very difficult due to the complexity of techniques like the classical shadow and 
tomography. Hence, better protocols with lower measurement complexity are still in demand.  

A feasible route for lower measurement complexity is to use a low-dimensional ancilla, especially a qubit ancilla. As long as the large 
system's QFI can be well transferred into the ancilla, the evaluation on the ancilla would avoid high complexity of the measurement. In the meantime, 
the existing protocols, such as those in the qubit scenario\,\cite{Yu2019,Tan2019,Macri2016,Zhang2026,Cerezo2021,Rath2021,Zhang2025,Yu2021,Yu2022}, 
can also be directly applied in the ancilla to evaluate the transferred QFI. However, it is still unknown whether this route is actually useful and, if so, when. 
Hence, the major questions in this field are (i) under what conditions the low-dimensional ancilla is useful for evaluating the QFI of the large system and 
(ii) how to construct the protocol when the ancilla is useful. These questions require systematic study, and addressing them is the main motivation for this 
work.

%================================ Figure ==================================
\begin{figure}[tp]
\centering
\includegraphics[width=\linewidth]{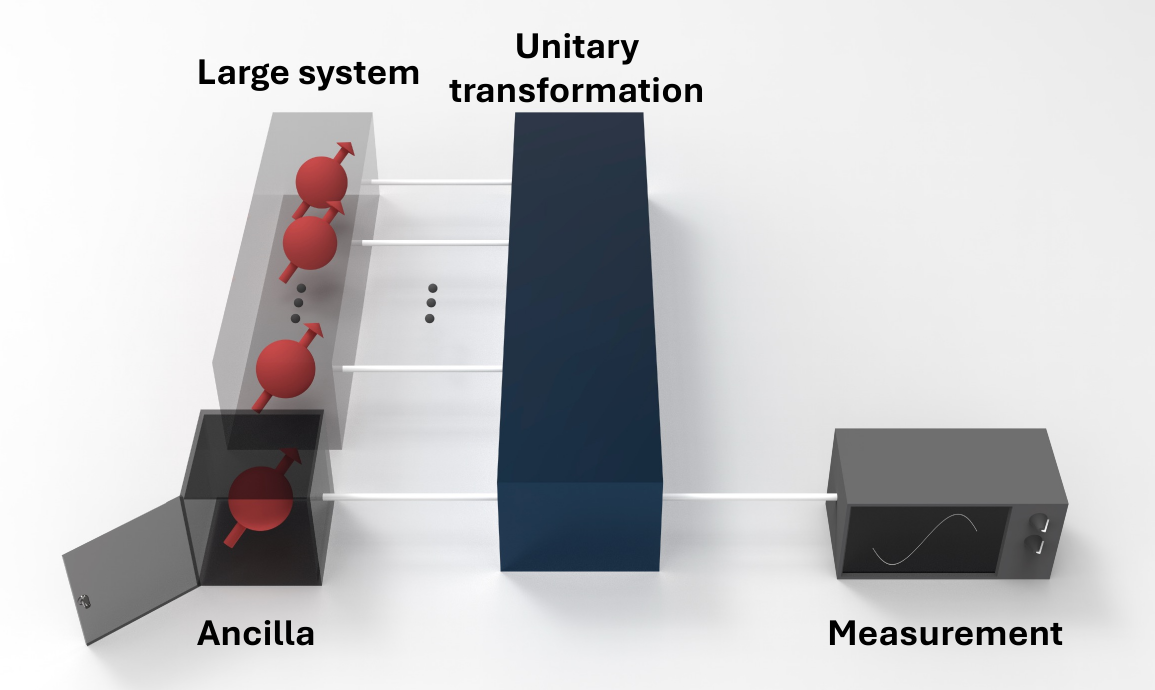} 
\caption{Demonstration of the evaluation protocol with an ancilla. Both the large system and 
ancilla are transformed by a unitary transformation, then the ancilla is measured to evaluate 
the QFI of the large system.}
\label{fig:schematic}
\end{figure}
%========================================================================

\emph{Optimal evaluation protocol.} 
The core of the protocol is to transfer the QFI of the large system (with dimension $N$) to an ancilla 
by a unitary transformation $U$ and then evaluate the ancilla's QFI via various methods. Denote the 
state of the composite system before the unitary transformation as $\rho(x)\otimes\rho_{a}$ with 
$\rho_{a}$ an $x$-independent state on the ancilla. The mission is to evaluate the QFI of $\rho(x)$ 
with respect to $x$. 

The first step of the protocol is to perform a unitary transformation $U=\sum_{i=1}^{N}\ket{i}\bra{i}\otimes T_{i}$ 
on both the system and ancilla, as demonstrated in Fig.\,\ref{fig:schematic}, where $\{\ketbra{i}\}$ is a set of 
projective measurements on the large system and $\{T_{i}\}$ is a set of unitary operators on the ancilla. Next, 
some lower bound of the QFI, such as the sub-QFI, is measured in the ancilla and the QFI of the large system 
is evaluated correspondingly. 

The unitary transformation $U$ and the measurement need to be designed for the best evaluation performance. 
The reduced density matrix of the ancilla after the transformation (denoted by $\tilde{\rho}_a$) reads 
$\tilde{\rho}_a=\sum^N_{i=1} p_i(x) T_i\rho_a T^{\dagger}_i$ with $p_{i}(x)=\bra{i}\rho(x)\ket{i}$. 
When the initial state of the ancilla is a pure state (denoted by $\ket{\phi}\bra{\phi}$), $\tilde{\rho}_a$ can be further 
expressed by 
\begin{equation}
\tilde{\rho}_a = \sum^N_{i=1} p_i(x) T_i\ket{\phi}\bra{\phi}T^{\dagger}_i. 
\label{eq:reduced_dm}
\end{equation}
Based on the property of the QFI, one can see that 
\begin{equation}
F(\rho(x)) = F(\rho(x)\otimes\rho_{a})\geq I(\{p_{i}\}), 
\end{equation}
where $I(\{p_{i}\})$ is the CFI of $\rho(x)$ with respect to the measurement $\{\ketbra{i}\}$. 
It is well-known that the inequality above can be saturated when $\{\ketbra{i}\}$ is optimal. An alternative choice of the optimal 
$\{\ketbra{i}\}$ is the eigenbasis of the SLD operator corresponding to $\rho(x)$.

For the quantum state in Eq.\,(\ref{eq:reduced_dm}), the corresponding QFI satisfies 
$F(\tilde{\rho}_a) \leq I(\{p_i\})+\sum_i p_i F(T_i\ket{\phi})$\,\cite{Alipour2015}. Here $F(T_i\ket{\phi})$ is the QFI for $T_i\ket{\phi}$. 
In this case, $T_i\ket{\phi}$ is independent of $x$, and $F(T_i\ket{\phi})$ then equals to zero. The proof of this inequality given in 
Ref.\,\cite{Alipour2015} is not fully correct and the rigorous proof is given in Appendix\,\ref{sec:apx_proof}. Furthermore, we have found 
the general attainability condition of this inequality, which is proved in Appendix\,\ref{sec:apx_A2}. The general attainability condition is 
given as the theorem below. 

\emph{Theorem 1.}
For the state $\tilde{\rho}=\sum^N_{i=1} p_i(x)\rho_i(x)$ with all $p_i(x)\neq 0$, the attainability condition for the inequality 
$F(\tilde{\rho}) \leq \sum^N_{i=1} p_i F(\rho_i) + I(\{p_i\})$ is that the eigenstate of $\rho_i$ within its support must also be the eigenstate 
of $L-L_i$ with the corresponding eigenvalue $\partial_x \ln p_i$, namely, 
\begin{equation}
\left(L - L_i\right)\ket{\lambda_{i,j}}=(\partial_x \ln p_i)\ket{\lambda_{i,j}},\,\forall \ket{\lambda_{i,j}}\in \mathrm{Supp}(\rho_i). 
\label{eq:theorem1_1}
\end{equation}
Here $L$ and $L_i$ are the SLD operators corresponding to $\tilde{\rho}$ and $\rho_i$, respectively. $\ket{\lambda_{i,j}}$ is the $j$th 
eigenstate of $\rho_i$ with respect to the eigenvalue $\lambda_{i,j}$. $\mathrm{Supp}(\rho_i)$ represents the support of $\rho_i$, i.e., 
the sub-eigenspace with respect to the nonzero eigenvalues. Equivalently, the condition above can also be expressed by 
\begin{equation}
\bra{\lambda_{i,l}}\left(L - L_i\right)\ket{\lambda_{i,j}}=(\partial_x \ln p_i)\delta_{jl},
\,\forall \ket{\lambda_{i,j(l)}}\in \mathrm{Supp}(\rho_i), 
\label{eq:theorem1_2}
\end{equation}
where $\delta_{jl}$ is the Kronecker delta function. 

In our protocol, the dimension of the ancilla (denoted by $n$) should be way smaller than that of the system. Let us 
further denote the set of distinct values of nonzero $\partial_x \ln p_i$ as $\{q_k\}^m_{k=1}$ (i.e., the underlying set of the multiset 
$\{\!\{\partial_x \ln p_i\,|\,\partial_x \ln p_i\neq 0, \forall i\}\!\}$) with $m$ the number of the distinct values. Under the condition 
that $\{\ketbra{i}\}$ is already optimal, we first present a no-go theorem based on Theorem 1. 

\emph{Theorem 2 (No-Go Theorem).} 
No ``perfect'' protocol exists for the evaluation of the system's QFI when $m>n$.

Here the ``perfect'' protocol refers to the one satisfying $F(\tilde{\rho}_a)=F(\rho)$ in theory. As a matter of fact, the number of 
Eq.\,(\ref{eq:theorem1_1}) is at most $n$ when $m>n$, indicating that there must exist some values of $\partial_x \ln p_i$ do not 
satisfy this equation. Hence, $F(\tilde{\rho}_a)$ cannot be equivalent to $F(\rho)$ in this case. 

When $m\leq n$, the condition in Eq.\,(\ref{eq:theorem1_2}) reduces to 
\begin{equation}
\left(\partial_x \ln p_i-\partial_x \ln p_j\right)\bra{\phi}T^{\dag}_i T_j\ket{\phi}=0,~\forall i,j.
\end{equation}
Detailed derivation of this equation can be found in Appendix\,\ref{sec:apx_attain_protocol}. When $\partial_x \ln p_i=\partial_x \ln p_j$, 
this equation is naturally satisfied. When $\partial_x \ln p_i\neq \partial_x \ln p_j$, this equation further reduces to 
$\bra{\phi}T^{\dag}_i T_j\ket{\phi}=0$, which means $T_i\ket{\phi}$ and $T_j\ket{\phi}$ have to be orthogonal. Hence, one can obtain 
the following theorem for the optimal $\{T_i\}$ that let $F(\tilde{\rho}_a)=F(\rho)$, where $\mathcal{T}_k$ is defined by 
$\mathcal{T}_k:=\{T_i\,|\,\partial_x\ln p_i = q_k\}$. 

\emph{Theorem 3.} 
In the case that $m\leq n$, the optimal $\{T_i\}$ must satisfy $\bra{\phi}T^{\dag}_i T_j\ket{\phi}=0$ for any $T_i$ and $T_j$ 
belonging to different sets of $\mathcal{T}_k$. A simple choice of the optimal $\{T_i\}$ is one that takes all $T_i$ belonging to 
the same $\mathcal{T}_k$ to be of the same form and in the meantime makes $\{T_i\ket{\phi}\}$ an orthonormal set of size $m$.  
 
When both $\{\ketbra{i}\}$ and $\{T_i\}$ are optimal, the left thing to construct an optimal protocol is to evaluate the QFI of 
the ancilla. In the case that the ancilla's QFI is still not accessible in practice, the lower bounds of the QFI, such as the sub-QFI 
$F_{\mathrm{sub}}(\tilde{\rho}_a)$\,\cite{Cerezo2021}, can be measured alternatively. The expression of the sub-QFI for pure 
states is given in Ref.\,\cite{Cerezo2021} and that for the mixed states is given in Appendix\,\ref{sec:apx_subQFI_mixed}. For 
pure states and any type of states in the two-level system, the sub-QFI is equivalent to the QFI (detailed derivations are given in 
Appendices\,\ref{sec:apx_subQFI_pure} and\,\ref{sec:apx_subQFI_2level}).  Therefore, for a qubit ancilla, the corresponding QFI 
can be evaluated by the tomography or the measurement methodology tailored to the sub-QFI like randomized 
measurement\,\cite{Yu2021}.

%================================ Figure ==================================
\begin{figure*}[tp]
\centering
\includegraphics[width=\linewidth]{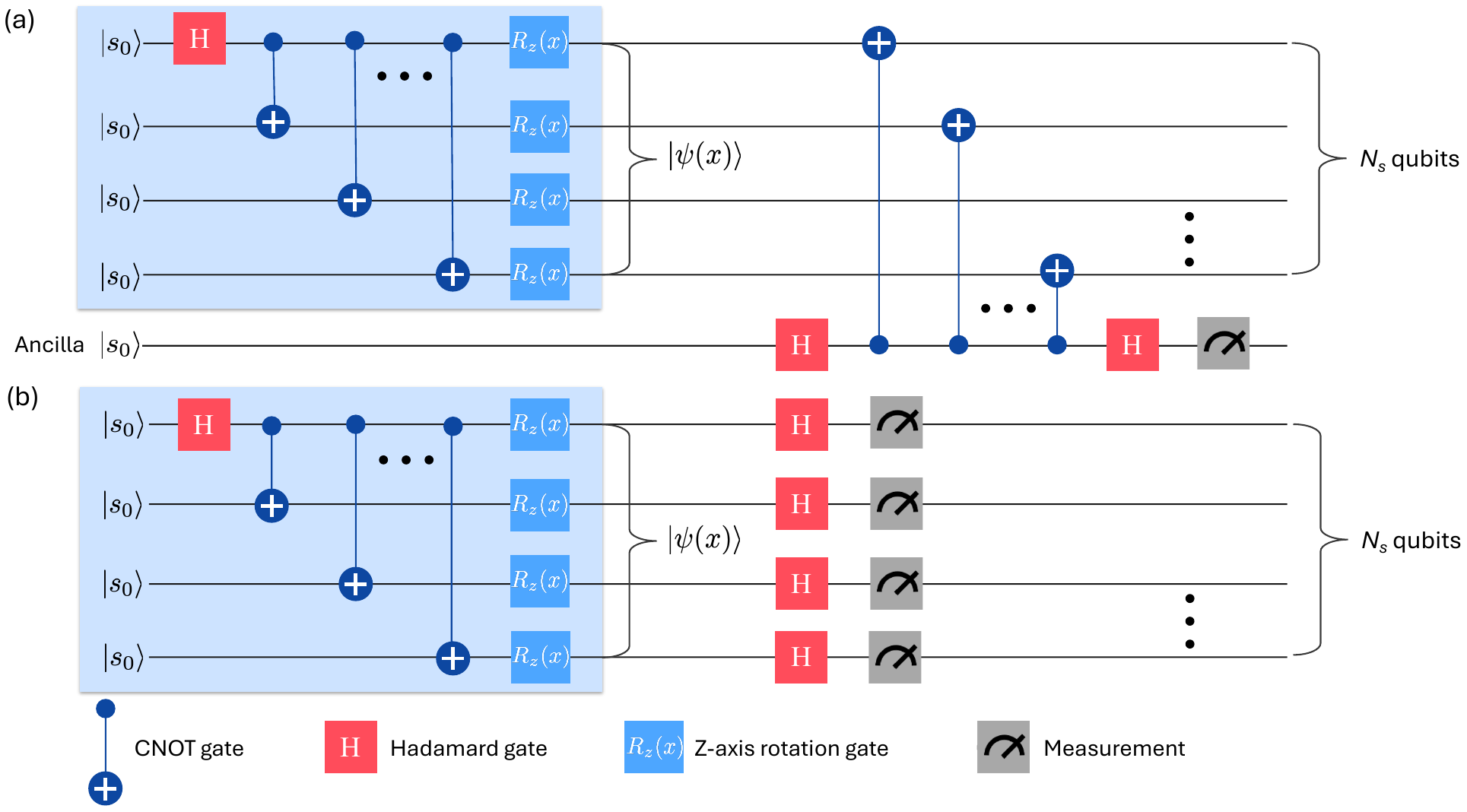} 
\caption{The circuit schematic to evaluate the QFI for the GHZ state. (a) The 
measurement via a qubit ancilla; (b) The direct measurement on the large system. 
The light-blue regions represent the preparation processes of the GHZ state. The 
measurement icon refers to the projective measurement $\{\ketbra{s_0}, \ketbra{s_1}\}^{\otimes N_s}$.}
\label{fig:circuit}
\end{figure*}
%========================================================================

An interesting case is that $\rho(x)$ is a pure state. In this case, we have the following corollary for the optimal protocol. 

\emph{Corollary 1.} 
For a pure $\rho(x)$, the optimal protocol constructed via a qubit ancilla can realize the best evaluation performance in theory, i.e., 
let $F_{\mathrm{sub}}(\tilde{\rho}_a)=F(\tilde{\rho}_a)=F(\rho)$. 

For a pure $\rho(x)$, when $\{\ketbra{i}\}$ is taken as the eigenbasis of the SLD operator corresponding to $\rho(x)$, $m$ must be $2$ 
due to the fact that the rank of this SLD operator is $2$ (see Appendix\,\ref{sec:qpx_rank} for the proof). Hence, the optimal $\{T_i\}$ 
must exist and can be taken as the simple one given in Theorem 3. More importantly, for a qubit ancilla the sub-QFI is equivalent to the 
QFI. Therefore, in theory the optimal protocol can realize $F_{\mathrm{sub}}(\tilde{\rho}_a)=F(\tilde{\rho}_a)=F(\rho)$, namely, the best 
performance in theory.  

\emph{Application in quantum circuits.}  
Next we use the platform of quantum circuits to show the entire evaluation protocol. Consider an $N_s$-qubit quantum circuit prepared in 
the Greenberger-Horne-Zeilinger (GHZ) state $\ket{\psi(x)} = (\ket{s_0}^{\otimes N_s} + e^{iN_s x}\ket{s_1}^{\otimes N_s})/\sqrt{2}$, where 
$\{\ket{s_0}, \ket{s_1}\}$ is the computational basis in the Hilbert space for one single qubit. Here the QFI of $\ket{\psi(x)}$ with respect 
to $x$ is required to be evaluated. The theoretical value of this QFI is $N_s^2$. 

In this example, the initial state of the ancilla is $\ket{s_0}$. A simple choice of the optimal 
$\{\ketbra{i}\}$ is $\{\ket{s_{+}}\bra{s_{+}}, \ket{s_{-}}\bra{s_{-}}\}^{\otimes N_s}$ with 
$\ket{s_{\pm}}=(\ket{s_0}\pm\ket{s_1})/\sqrt{2}$. Utilizing this measurement, the conditional probability for obtaining $k$ occurrences of 
$s_{-}$ in the outcome is 
\begin{equation}
p_k=\frac{1}{2^{N_s}}\binom{N_s}{k} [1+(-1)^{k}\cos(N_sx)], 
\end{equation}
where $\binom{N_s}{k}$ is the binomial coefficient. Then the corresponding CFI is  
\begin{equation}
I=\sum_{k=0}^{N_s} \frac{(\partial_{x} p_k)^2}{p_k}= N_s^2,
\label{eq:CFI_circuit}
\end{equation}
which is obviously equivalent to the QFI, indicating that $\{\ket{s_{+}}\bra{s_{+}}, \ket{s_{-}}\bra{s_{-}}\}^{\otimes N_s}$ is an optimal 
choice of $\{\ketbra{i}\}$. Next, based on Theorem 3 we can construct the operator $U$ as follows
\begin{equation}
U=\sum_{l_1,\cdots,l_{N_s}=\pm} \ket{s_{l_1}\cdots s_{l_{N_s}}}\bra{s_{l_1}\cdots s_{l_{N_s}}}\otimes T_{l_1,\cdots,l_{N_s}}. 
\label{eq:opt_U}
\end{equation}
Here $T_{l_1,\cdots,l_{N_s}}$ for the even (denoted by $T_{\mathrm{even}}$) and odd $k$ (denoted by $T_{\mathrm{odd}}$) can be taken as 
any two matrices satisfying $\bra{s_0}T^{\dag}_{\mathrm{even}}T_{\mathrm{odd}}\ket{s_0}=0$ based on Theorem 3. For the sake of providing 
a simple scheme, we take $T_{\mathrm{even}}=\openone_2$ and $T_{\mathrm{odd}}=\sigma_x$. $\openone_2$ is the two-dimensional identity 
operator and $\sigma_x=\ket{s_0}\bra{s_1}+\ket{s_1}\bra{s_0}$ is the Pauli X operator. The two-qubit (Controlled-NOT) CNOT gate reads 
$C=\ket{s_0}\bra{s_0}\otimes\openone_2+\ket{s_1}\bra{s_1}\otimes\sigma_x$. As a matter of fact, the operation in Eq.\,(\ref{eq:opt_U}) can be 
decomposed as 
\begin{equation}
U=\left(H^{\otimes N_s}\otimes\openone_2\right) \left(\prod_{i=1}^{N_s}  C_i \right) \left(H^{\otimes N_s}\otimes\openone_2\right),
\label{eq:U_c}
\end{equation}
where $C_i$ is the CNOT gate on each qubit and the ancilla, i.e., 
$C_1=\ket{s_{0}}\bra{s_{0}}\otimes \openone^{\otimes N_s}_2+\ket{s_{1}}\bra{s_{1}}\otimes \openone^{\otimes (N_s-1)}_2\otimes\sigma_x$, and 
\begin{align}
C_i=&\openone^{\otimes (i-1)}_2\otimes \ket{s_{0}}\bra{s_{0}}\otimes \openone^{\otimes (N_s-i+1)}_2 \nonumber \\
& +\openone^{\otimes (i-1)}_2\otimes \ket{s_{1}}\bra{s_{1}}\otimes \openone^{\otimes (N_s-i)}_2\otimes\sigma_x
\end{align}
for $i=2,3,\cdots,N_s-1$, and $C_{N_s}=\openone^{\otimes (N_s-1)}_2\otimes \ket{s_{0}}\bra{s_{0}} \otimes\openone_2
+ \openone^{\otimes (N_s-1)}_2\otimes\ket{s_{1}}\bra{s_{1}} \otimes\sigma_x$. $H$ is the Hadamard gate for a single qubit. In the basis 
$\{\ket{s_0},\ket{s_1}\}$, it reads 
\begin{equation}
H = \frac{1}{\sqrt{2}}
\begin{pmatrix}
1 & 1 \\
1 & -1
\end{pmatrix}. 
\end{equation}

The operation in Eq.\,(\ref{eq:U_c}) requires the existence of the Hadamard gate on each qubit. The complexity of the circuit can be further 
reduced by utilizing the reversed CNOT gate, which is defined by  $C_{\mathrm{rev},i}\!=\!\openone^{\otimes N_s}_2\otimes \ket{s_{0}}
\bra{s_{0}}+\openone^{\otimes (i-1)}_2\otimes \sigma_x\otimes \openone^{\otimes (N_s-i)}_2\otimes\ket{s_{1}}\bra{s_{1}}$. 
Its relation with the CNOT gate can be expressed by $C_i=H^{\otimes (N_s+1)} C_{\mathrm{rev},i} H^{\otimes (N_s+1)}$. 
Utilizing this relation, Eq.\,(\ref{eq:U_c}) reduces to 
\begin{equation}
U=\left(\openone^{\otimes N_s}_2\otimes H\right) \left(\prod_{i=1}^{N_s}  C_{\mathrm{rev},i} \right) 
\left(\openone^{\otimes N_s}_2\otimes H\right). 
\end{equation}
The quantum circuit to realize this operation is given in Fig.\,\ref{fig:circuit}(a). The light-blue region in the circuit represents the preparation 
process of the GHZ state. In the basis $\{\ket{s_0}, \ket{s_1}\}$, the Z-axis rotation gate reads 
\begin{equation}
R_z(x) = \begin{pmatrix}
e^{-i\frac{x}{2}} & 0 \\
0 & e^{i\frac{x}{2}}
\end{pmatrix}. 
\end{equation}
After the operation of $U$, the QFI of the qubit ancilla needs to be evaluated. In this case, the density matrix of the ancilla is diagonal and the 
tomography can be efficiently applied to evaluate the entries of the density matrix, and the corresponding derivatives via tuning the parameter 
$x$ to $x+\mathrm{d}x$, or $x-\pi/(2N_s)$ to avoid the finite-difference method. 

In this example, the QFI  in the large system can be evaluated via the measurement of sub-QFI using randomized measurements, as 
illustrated by Yu et al.~\cite{Yu2021} in the 4-qubit case. In the meantime, it can also be directly measured by performing the 
Hadamard gate on each qubit and then taking the projective measurement $\{\ketbra{s_0}, \ketbra{s_1}\}^{\otimes N_s}$, as shown in 
Fig.\,\ref{fig:circuit}(b). Multiple rounds of this entire measurement process have to be performed to evaluate the set $\{p_k\}$. Next, 
the gate $R_z(x)$ should be replaced by $R_z(x+\mathrm{d}x)$ and the entire measurement process also needs to be repeated multiple 
rounds for the sake of evaluating $\{\partial_x p_k\}$. Finally, the CFI is evaluated via Eq.\,(\ref{eq:CFI_circuit}) and the value of QFI is 
then obtained since $\{\ketbra{s_{+}}, \ketbra{s_{-}}\}^{\otimes N_s}$ is optimal.  In both ways the measurement complexity would grow at 
least polynomially~\cite{Huang2020,Hu2023} as the increase with $N_s$.  

On the other hand, at the current state of the art, the fidelities of single- and two-qubit gates are all higher than 99\%\,\cite{McKay2017,
Spring2022,Somoroff2023,Sung2021}, yet the faithful rate of the qubit reading is only $96\% \sim 99\%$\,\cite{Abdurakhimov2024,Lambert2025}. 
Hence, the evaluation scheme in Fig.\,\ref{fig:circuit}(b) would be strongly affected by the readout error of the qubits, especially the inconsistency 
of this error for different qubits. Therefore, this direct measurement scheme would not provide a high-precision evaluation performance. 

In the ancilla-assisted protocol shown in Fig.\,\ref{fig:circuit}(a), it is not very difficult to choose a good qubit with a low readout error in 
practice, and due to the fact that the measurement is only performed on the ancilla, the inconsistency of this error among different 
qubits is not a problem anymore. The major difficulty to realize this protocol is that it requires the circuit depth of $N_s$, and a deep   
circuit has to face additional noises, and we admit that currently the corresponding performance may not be very ideal for a large $N_s$. 
However, we also believe that this noise would be reduced in the near future with further development in quantum information,  yet the 
measurement complexity in the scheme in Fig.\,\ref{fig:circuit}(b) cannot be reduced. Therefore, our protocol will be more advantageous 
in the future and is highly likely to become the mainstream approach for QFI measurement in this system.

\emph{Conclusion and outlook.} 
In this paper we have answered the questions (i) under what conditions the low-dimensional ancilla is useful for evaluating the QFI of the 
large system and (ii) how to construct the protocol when the ancilla is useful. Utilizing the attainability condition given in Theorem 1, we 
show that no ``perfect'' protocol exists when the number of distinct values in the multiset $\{\!\{ p_i\}\!\}^N_{i=1}$ is larger than the 
dimension of the ancilla, as stated in Theorem 2. In the case that this number is no larger than the ancilla's dimension, the condition for 
the optimal scheme is thoroughly provided and discussed. The protocol has been applied into the system of $N_s$-qubit quantum circuits 
as an illustration to evaluate the QFI of the GHZ state. 

This work starts the systematic research on the route of evaluating the QFI of large systems via the ancilla, and we believe that it will 
motivate further experimental studies on the evaluation of the QFI. Meanwhile, in scenarios where the QFI of the large system cannot 
be ``perfectly'' transferred to the ancilla, many interesting open questions remain to be further investigated, including how to construct 
an optimal scheme to achieve the best evaluation performance. We believe that as these problems are resolved, the precision of QFI 
evaluation will continue to improve, which will greatly advance the development of fields such as the detection of entanglement and 
phase transitions.

\begin{acknowledgments}
This work was supported by the National Natural Science Foundation of China (Grants No.\,12575013 and No.\,12547103).  
DeepSeek-V4.1-Flash was used to polish the English writing of the paper.   
\end{acknowledgments}

\appendix

\section{Maximization of the QFI}

\subsection{Proof of the inequality}
\label{sec:apx_proof}

Consider a quantum state in the form 
\begin{equation}
\rho = \sum^N_{i=1} p_i(x) \rho_i(x),
\end{equation}
 where $\rho_i(x)$ is a density matrix relies on $x$, and $p_i(x)$ is a weight relies on $x$ and satisfies $\sum^N_{i=1} p_i=1$. 
Notice that in the main text the symbol $\tilde{\rho}$ is used to denote $\sum^N_{i=1} p_i(x) \rho_i(x)$. Here we use $\rho$ instead 
for the sake of a neat presentation. Without loss of generality, we can require $p_i(x)\neq 0$. The quantum Fisher information (QFI) 
of this state satisfies the following inequality
\begin{equation}
F(\rho) \leq \sum^N_{i=1} p_i F(\rho_i) + I(\{p_i\}),
\label{eq:apx_ineq}
\end{equation}
where $F(\rho_i)$ is the QFI of the state $\rho_i$ and $I(\{p_i\})$ is the CFI of the distribution $\{p_i\}$, i.e., 
\begin{equation}
I(\{p_i\})=\sum^N_{i=1} \frac{(\partial_x p_i)^2}{p_i}.
\end{equation}
The inequality (\ref{eq:apx_ineq}) was first given in Ref.\,\cite{Alipour2015}. This inequality is correct, however, the proof given in 
Ref.\,\cite{Alipour2015} is not correct. We first show why their proof is wrong and then provide the right proof of this inequality. 

In the derivation of Eq.\,(A3) in the Appendix A in Ref.\,\cite{Alipour2015}, the inequality 
\begin{equation}
\sum^m_{i=1} \left| \frac{\sum^N_{j=1} p_j\Tr(\Pi_i \mathsf{L}_j \rho_j)}{\sqrt{\Tr(\Pi_i \rho)}}\right|^2
\leq \sum^m_{i=1}\sum^N_{j=1} \left| \frac{p_j\Tr(\Pi_i \mathsf{L}_j \rho_j)}{\sqrt{\Tr(\Pi_i \rho)}}\right|^2
\label{eq:apx_difference}
\end{equation}
is used, where $\Pi_i$ is an element of a POVM $\{\Pi_i\}^m_{i=1}$ and $\mathsf{L}_j$ satisfies 
\begin{equation}
\partial_x (p_j\rho_j)=\frac{1}{2}p_j(\mathsf{L}_j \rho_j+\rho_j \mathsf{L}_j).
\label{eq:apx_L}
\end{equation}
Unfortunately, the inequality (\ref{eq:apx_difference}) does not hold in general. As a matter of fact, for a group of complex numbers 
$\{z_i\}$ the inequality $\left|\sum_i z_i\right|^2 \leq \sum_i |z_i|^2$ does not generally hold. The true triangle inequality reads 
$\left|\sum_i z_i\right| \leq \sum_i |z_i|$ or $\left|\sum_i z_i\right|^2 \leq (\sum_i |z_i|)^2$. In the following we provide a counterexample 
of Eq.\,(\ref{eq:apx_difference}). 

\begin{figure}[tp]
\centering\includegraphics[width=\linewidth]{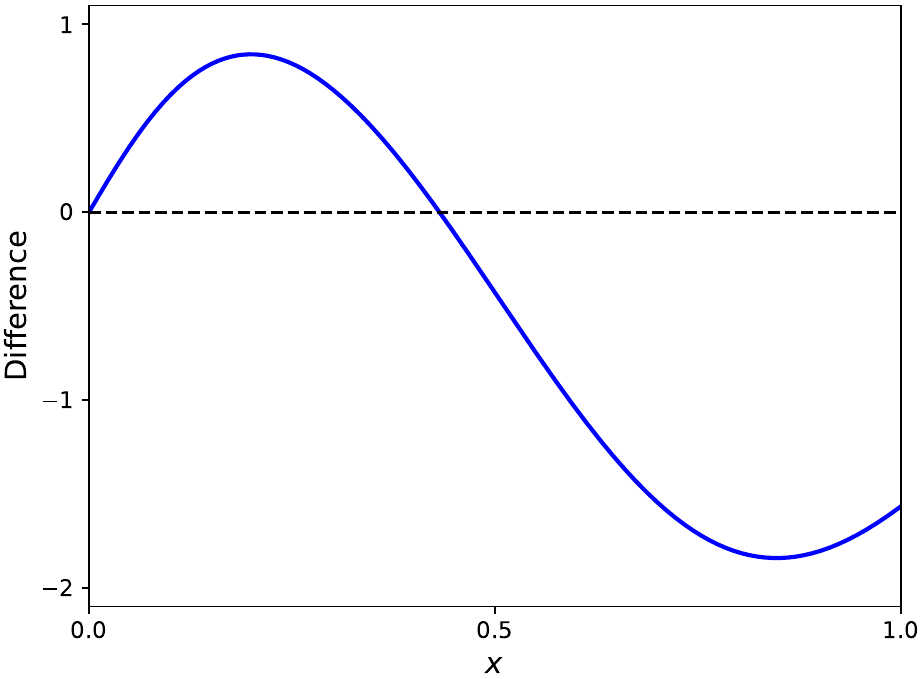} 
\caption{The difference between $\sum^m_{i=1} \left| \frac{\sum^N_{j=1} p_j\Tr(\Pi_i \mathsf{L}_j \rho_j)}{\sqrt{\Tr(\Pi_i \rho)}}\right|^2$ 
and $\sum^m_{i=1}\sum^N_{j=1} \left| \frac{p_j\Tr(\Pi_i \mathsf{L}_j \rho_j)}{\sqrt{\Tr(\Pi_i \rho)}}\right|^2$ as a function of $x$ in the 
demonstrated example. }
\label{fig:apx_inequality}
\end{figure}

Consider the quantum state 
\begin{equation}
\rho = \sin^2(x)\rho_1+\cos^2(x)\rho_2
\end{equation}
with the expressions of $\rho_1$ and $\rho_2$ are
\begin{eqnarray}
\rho_1 &=& \frac{1}{2}\mqty(
1 & 1 \\ 
1 & 1), \\
\rho_2 &=& \mqty(
\cos^2(x) & \cos(x)\sin(x) e^i \\ 
\cos(x)\sin(x) e^{-i} &\sin^2(x)).
\end{eqnarray}
Here the basis of the matrices is $\{\ket{0}, \ket{1}\}$ (the computational basis for a qubit). The positive operator valued measure (POVM) 
$\{\Pi_1, \Pi_2, \Pi_3\}$ is taken as 
\begin{eqnarray}
\Pi_1 &=& \frac{\sqrt{2}}{1+\sqrt{2}}\ket{0}\bra{0}, \\
\Pi_2 &=& \frac{\sqrt{2}}{2(1+\sqrt{2})}(\ket{0}-\ket{1})(\bra{0}-\bra{1}),
\end{eqnarray}
and $\Pi_3 = \openone-\Pi_1-\Pi_2$. The difference between $\sum^m_{i=1} \left| \frac{\sum^N_{j=1} p_j\Tr(\Pi_i \mathsf{L}_j \rho_j)}
{\sqrt{\Tr(\Pi_i \rho)}}\right|^2$ and $\sum^m_{i=1}\sum^N_{j=1} \left| \frac{p_j\Tr(\Pi_i \mathsf{L}_j \rho_j)}{\sqrt{\Tr(\Pi_i \rho)}}\right|^2$ 
for different values of $x$ is shown in Fig.\,\ref{fig:apx_inequality}. It can be seen that this difference is not always negative, which means 
that the inequality in Eq. (\ref{eq:apx_difference}) does not always hold. 

Furthermore, if the correct inequality 
\begin{equation}
\sum^m_{i=1} \left| \frac{\sum^N_{j=1} p_j\Tr(\Pi_i \mathsf{L}_j \rho_j)}{\sqrt{\Tr(\Pi_i \rho)}}\right|^2
\leq \sum^m_{i=1}\!\frac{\left(\sum^N_{j=1} | p_j\Tr(\Pi_i \mathsf{L}_j\rho_j)|\right)^2}{\Tr(\Pi_i \rho)}
\end{equation}
is applied, the final inequality becomes
\begin{equation}
F(\rho) \leq N\left[\sum^N_{i=1} p_i F(\rho_i) +  I(\{p_i\})\right].
\label{eq:apx_inequl2}
\end{equation}
This inequality is way looser than that given in Eq.\,(\ref{eq:apx_ineq}). 

%=========================================== Figure ================================================
%=================================================================================================

Although the proof given in Ref.\,\cite{Alipour2015} is not correct, the inequality in Eq.\,(\ref{eq:apx_ineq}) is correct. Next we provide 
the rigorous proof of this inequality. Consider a quantum state in the form 
\begin{equation}
\rho_{\mathrm{tot}}=\sum^N_{i=1}p_i\rho_i\otimes \ket{i}\bra{i},
\label{eq:apx_rhotot0}
\end{equation}
where $\otimes$ represents the tensor product and $\{\ket{i}\}$ is an orthogonal basis in the $N$-dimensional Hilbert space. 
Expand this state in the second Hilbert space with the basis $\{\ket{i}\}$, it can also be written as 
\begin{equation}
\rho_{\mathrm{tot}}=\bigoplus^N_{i=1} p_i \rho_i,
\label{eq:apx_rhotot1}
\end{equation}
where $\oplus$ represents the direct sum. The symmetric logarithmic derivative (SLD) operator for $\rho_{\mathrm{tot}}$ can be 
expressed by\,\cite{Liu2014,Liu2016}
\begin{equation}
L_{\mathrm{tot}}=\bigoplus^N_{i=1} \mathsf{L}_i, 
\label{eq:apx_Ltot1}
\end{equation}
where $\mathsf{L}_i$ satisfies Eq.\,(\ref{eq:apx_L}) with respect to $p_i \rho_i$. As a matter of fact,  $\mathsf{L}_i$ can be further 
expressed by 
\begin{equation}
\mathsf{L}_i=\frac{\partial_x p_i}{p_i} + L_{i}
\label{eq:apx_Li}
\end{equation}
with $L_i$ the SLD operator for $\rho_i$. Then the QFI of $\rho_{\mathrm{tot}}$ can be written as  
\begin{equation}
F(\rho_{\mathrm{tot}}) = \Tr(\rho_{\mathrm{tot}} L^2_{\mathrm{tot}})=\sum_i \Tr(p_i \rho_i \mathsf{L}^2_i).
\end{equation}
Utilizing Eq.\,(\ref{eq:apx_Li}), one can see that 
\begin{eqnarray}
F(\rho_{\mathrm{tot}}) &=& \sum^N_{i=1}p_i\Tr( \rho_i L_i^2)+\frac{(\partial_x p_i)^2}{p_i} \nonumber \\
&=& \sum^N_{i=1} p_i F(\rho_i) + I(\{p_i\}).
\label{eq:apx_QFI}
\end{eqnarray}

Furthermore, due to the fact that the QFI is monotonic under the completely positive and trace-preserving map,  one can further obtain 
\begin{equation}
F(\rho) \leq F(\rho_{\mathrm{tot}}), 
\end{equation}
namely, 
\begin{equation}
F(\rho) \leq \sum^N_{i=1} p_i F(\rho_i) + I(\{p_i\}).
\end{equation}
The inequality is then proved. 

Compared this result with Eq.\,(\ref{eq:apx_inequl2}), it is easy to see that the inequality above is always tighter than Eq.\,(\ref{eq:apx_inequl2}). 
Hence, Eq.\,(\ref{eq:apx_inequl2}) is basically useless in practice. 

\subsection{Attainability of the inequality}
\label{sec:apx_A2}

In the following we provide the proof of the necessary and sufficient conditions to attain the inequality\,(\ref{eq:apx_ineq}). 
For the state in Eq.\,(\ref{eq:apx_rhotot0}), the operation of partial trace on the second Hilbert space can be represented by a set 
of Kraus operators $\{K_i\}$ with 
\begin{equation}
K_i  = \openone\otimes \bra{i},
\end{equation}
where $\openone$ is the identity operator. It is easy to check that  
\begin{equation}
\sum^N_{i=1}K_i\rho_{\mathrm{tot}}K^{\dag}_i=\sum^N_{i=1}p_i\rho_i=\rho.
\end{equation} 
Hence, the QFI for $\sum^N_{i=1}K_i\rho_{\mathrm{tot}}K^{\dag}_i$ is equivalent to that for $\rho$. Next we calculate this QFI.  
Denote the SLD with respect to the state  $\rho$ as $L$, it is easy to see that 
\begin{align}
F(\rho )  &= \Tr(\sum^N_{i=1} K_i \rho_{\mathrm{tot}} K_i^\dagger L^2) \nonumber \\
&=\Tr(\rho_{\mathrm{tot}}\sum^N_{i=1} K_i^{\dag} L^2 K_i). 
\end{align}
Utilizing the expression $K_i=\openone\otimes\bra{i}$, one can see that 
\begin{align}
\sum^N_{i=1} K_i^{\dag} L^2 K_i &=\sum^N_{i=1} L^2\otimes\ket{i}\bra{i} \nonumber \\
&= L^2\otimes \openone \nonumber \\
&= (L\otimes \openone)^2.
\end{align}
Hence, one can further obtain 
\begin{equation}
F(\rho) =\Tr(\rho_{\mathrm{tot}} (L\otimes \openone)^2). 
\end{equation}
Next we calculate the difference between $F(\rho_{\mathrm{tot}})$ and $F(\rho)$, which reads 
\begin{widetext}
\begin{eqnarray}
& & F(\rho_{\mathrm{tot}}) - F(\rho) \nonumber \\ 
&=& \Tr(\rho_{\mathrm{tot}} L_{\mathrm{tot}}^2) - \Tr(\rho_{\mathrm{tot}} (L\otimes \openone)^2) \nonumber \\
&=& \Tr(\rho_{\mathrm{tot}} \left(L_{\mathrm{tot}}-L\otimes \openone\right)^2)
+\Tr((\rho_{\mathrm{tot}} L_{\mathrm{tot}}+L_{\mathrm{tot}}\rho_{\mathrm{tot}})(L\otimes \openone))
-2\Tr(\rho_{\mathrm{tot}} (L\otimes \openone)^2) \nonumber \\
&=& \Tr(\rho_{\mathrm{tot}} (L_{\mathrm{tot}}-L\otimes \openone)^2)+2\Tr((\partial_x \rho_{\mathrm{tot}}) (L\otimes \openone))
-2\Tr(\rho_{\mathrm{tot}} (L\otimes \openone)^2). 
\label{eq:apx_temp1}
\end{eqnarray}
Notice that 
\begin{equation}
\sum^N_{i=1} K_i^{\dag} L K_i =\sum^N_{i=1} L\otimes\ket{i}\bra{i} = L\otimes \openone,
\end{equation}
then one can have 
\begin{eqnarray}
& & 2\Tr((\partial_x \rho_{\mathrm{tot}}) (L\otimes \openone))-2\Tr(\rho_{\mathrm{tot}} (L\otimes \openone)^2) \nonumber \\ 
&=& 2\Tr(\sum^N_{i=1} (\partial_x \rho_{\mathrm{tot}}) K_i^{\dag} L K_i )-2\Tr(\rho_{\mathrm{tot}}\sum_i K^{\dag}_i L^2 K_i) \nonumber \\
&=& 2\Tr(\left[\partial_x \left(\sum^N_{i=1} K_i \rho_{\mathrm{tot}}  K_i^{\dag}\right)\right] L)
-2\Tr( \left(\sum^N_{i=1} K_i \rho_{\mathrm{tot}} K^{\dag}_i \right) L^2 ) \nonumber \\
&=& 2\Tr(\left(\partial_x \rho\right) L)-2\Tr(\rho L^2 ) \nonumber \\
&=& 0.
\label{eq:apx_temp2}
\end{eqnarray}
\end{widetext}
Combining Eqs.\,(\ref{eq:apx_temp1}) and (\ref{eq:apx_temp2}) we finally obtain that 
\begin{equation}
F(\rho_{\mathrm{tot}}) - F(\rho)=\Tr(\rho_{\mathrm{tot}} (L_{\mathrm{tot}}-L\otimes \openone)^2) \geq 0.
\end{equation}
Hence, the attainability condition becomes 
\begin{equation}
\Tr(\rho_{\mathrm{tot}} (L_{\mathrm{tot}}-L\otimes \openone)^2)=0.
\label{eq:apx_attaincond}
\end{equation}
In the case that $\rho_{\mathrm{tot}}$ is full-rank, this condition reduces to $L_{\mathrm{tot}} = L\otimes \openone$. 
Noticing that $L_{\mathrm{tot}}$ can be written as 
\begin{equation}
L_{\mathrm{tot}} = \sum_{i=1}^{N} \left(\frac{\partial_x p_i}{p_i}+L_i\right)\otimes\ket{i}\bra{i}, 
\end{equation}
the condition above can be further expressed by 
\begin{equation}
\frac{\partial_x p_i}{p_i}+L_i = \frac{\partial_x p_j}{p_j}+L_j,~\forall i,j. 
\end{equation}

As a matter of fact, in many cases $\rho_{\mathrm{tot}}$ is not full-rank. Next we discuss the equivalent form of 
Eq.\,(\ref{eq:apx_attaincond}) for a general $\rho_{\mathrm{tot}}$. Denote the spectral decomposition of $\rho_i$ as 
\begin{equation}
\rho_i = \sum_j \lambda_{i,j} \ket{\lambda_{i,j}}\bra{\lambda_{i,j}},
\end{equation}
where $\lambda_{i,j}\geq 0$ is the $j$th eigenvalue of $\rho_i$ and $\ket{\lambda_{i,j}}$ is the corresponding eigenstate. 
Based on Eq.\,(\ref{eq:apx_rhotot0}), one can have 
\begin{equation}
\rho_{\mathrm{tot}} = \sum_{i,j} p_i \lambda_{i,j} \ket{\lambda_{i,j}}\bra{\lambda_{i,j}} \otimes \ket{i}\bra{i}.
\end{equation}
Then 
\begin{eqnarray*}
& &\Tr\left(\rho_{\mathrm{tot}}(L_{\mathrm{tot}}-L\otimes \openone)^2\right) \\
&=&\sum_{i,j} p_i \lambda_{i,j} \bra{\lambda_{i,j}}\otimes\bra{i}(L_{\mathrm{tot}}-L\otimes \openone)^2
\ket{\lambda_{i,j}}\otimes\ket{i}.
\end{eqnarray*}
Recall that we have required $p_i\neq 0$ without loss of generality. To fulfill the attainability condition in Eq.\,(\ref{eq:apx_attaincond}), 
for all the indices $i,j$ satisfying $\lambda_{i,j}\neq 0$ one must have 
\begin{equation}
\bra{\lambda_{i,j}}\otimes\bra{i}(L_{\mathrm{tot}}-L\otimes \openone)^2\ket{\lambda_{i,j}}\otimes\ket{i} = 0.
\end{equation}
Since $L_{\mathrm{tot}}-L\otimes \openone$ is a Hermitian operator, the equation above is equivalent to 
\begin{equation}
(L_{\mathrm{tot}}-L\otimes \openone)\ket{\lambda_{i,j}}\otimes\ket{i}=0.
\label{eq:apx_temp_cond1}
\end{equation}
Utilizing the expression 
\begin{equation}
L_{\mathrm{tot}}-L\otimes\openone = \sum_{k=1}^{N} \left(\frac{\partial_x p_k}{p_k}+L_k-L\right)\otimes\ket{k}\bra{k},
\end{equation}
Eq.\,(\ref{eq:apx_temp_cond1}) can further reduce to  
\begin{equation}
\left(\frac{\partial_x p_i}{p_i}+L_i - L\right)\ket{\lambda_{i,j}}=0,
\end{equation}
namely, 
\begin{equation}
\left(L - L_i \right)\ket{\lambda_{i,j}}=\frac{\partial_x p_i}{p_i}\ket{\lambda_{i,j}}.
\end{equation}

Denote $\mathrm{Supp}(\rho_i)$ as the support of $\rho_i$, where the eigenstate $\ket{\lambda_{i,j}}$ corresponds to the nonzero eigenvalue 
$\lambda_{i,j}$. The result above means that the necessary and sufficient condition to attain $F(\rho_{\mathrm{tot}})=F(\rho)$ is 
\begin{equation}
\left(L - L_i\right)\ket{\lambda_{i,j}}=\frac{\partial_x p_i}{p_i}\ket{\lambda_{i,j}},~\forall \ket{\lambda_{i,j}} \in \mathrm{Supp}(\rho_i),
\end{equation}
namely, the eigenstate of $\rho_i$ within its support must also be the eigenstate of $L-L_i$, with the corresponding 
eigenvalue $\partial_x p_i/p_i$. This equation can be further expressed by 
\begin{equation}
\bra{\lambda_{i,l}}\left(L - L_i\right)\ket{\lambda_{i,j}}=\frac{\partial_x p_i}{p_i}\delta_{jl},
~\forall \ket{\lambda_{i,j(l)}}\in \mathrm{Supp}(\rho_i). 
\end{equation}
Here $\delta_{jl}$ is the Kronecker delta function. 

\subsection{Attainability condition in the protocol}
\label{sec:apx_attain_protocol}

Recall that in our protocol, $\rho_i=T_i\ket{\phi}\bra{\phi}T^{\dag}_i$ is independent of $x$, then the attainability condition reduces to 
\begin{equation}
\bra{\phi}T^{\dag}_j L T_i\ket{\phi}=(\partial_x \ln p_i)\bra{\phi}T^{\dag}_j T_i\ket{\phi},~\forall i,j. 
\label{eq:apx_attainTi}
\end{equation}
Exchange the indices $i$ and $j$, it becomes 
\begin{equation}
\bra{\phi}T^{\dag}_i L T_j\ket{\phi}=(\partial_x \ln p_j)\bra{\phi}T^{\dag}_i T_j\ket{\phi},~\forall i,j. 
\label{eq:apx_attain_temp1}
\end{equation}
In the meantime, taking the complex conjugate of Eq.\,(\ref{eq:apx_attainTi}) and noticing that $L$ is Hermitian, 
one can have 
\begin{equation}
\bra{\phi}T^{\dag}_i L T_j\ket{\phi}=(\partial_x \ln p_i)\bra{\phi}T^{\dag}_i T_j\ket{\phi},~\forall i,j. 
\label{eq:apx_attain_temp2}
\end{equation}
Equations\,(\ref{eq:apx_attain_temp1}) and (\ref{eq:apx_attain_temp2}) indicate that 
\begin{equation}
\left(\partial_x \ln p_i-\partial_x \ln p_j\right)\bra{\phi}T^{\dag}_i T_j\ket{\phi}=0,~\forall i,j.
\end{equation}
When $\partial_x \ln p_i=\partial_x \ln p_j$, this equation is naturally satisfied. When $\partial_x \ln p_i\neq \partial_x \ln p_j$, 
this equation reduces to 
\begin{equation}
\bra{\phi}T^{\dag}_i T_j\ket{\phi}=0.
\end{equation}
namely, $T_i\ket{\phi}$ and $T_j\ket{\phi}$ are orthogonal. 

\section{Derivation of the sub-QFI}

\subsection{Derivation}
\label{sec:apx_subQFI_mixed}

The superfidelity is defined by\,\cite{Miszczak2009} 
\begin{equation}
g(\rho_{1},\rho_{2})=\mathrm{Tr}(\rho_{1}\rho_{2})
+\sqrt{\left[1-\mathrm{Tr}(\rho_{1}^{2})\right]\left[1-\mathrm{Tr}(\rho_{2}^{2})\right]},
\end{equation}
where $\rho_1$ and $\rho_2$ are two quantum states. Similar to the relation between QFI and fidelity, the second-order term of 
the superfidelity is naturally a lower bound of the QFI. Hence, it can be called the sub-QFI. Cerezo et al.\,\cite{Cerezo2021} 
have provided the expression of the sub-QFI for unitary parameterization processes. Here we provide the derivation of the sub-QFI 
for a general parameterization process. 

The sub-QFI is defined as 
\begin{equation}
F_{\mathrm{sub}}:=8\lim_{\mathrm{d}x\to0}\frac{1-\sqrt{g\left(\rho_x,\rho_{x+\mathrm{d}x}\right)}}
{\left(\mathrm{d}x\right)^{2}}. 
\label{eq:app_sub-qfi}
\end{equation}
When $\rho_x$ is a pure state, $g(\rho_x,\rho_{x+\mathrm{d}x})$ reduces to the fidelity, and the sub-QFI is nothing but the QFI 
itself. Next we discuss the case of mixed states. Utilizing the Taylor expansion up to the second order
\begin{equation}
\rho_{x+\mathrm{d}x}=\rho_x+\partial_{x}\rho_x\mathrm{d}x
+\frac{1}{2}\partial_{x}^{2}\rho_x(\mathrm{d}x)^{2}
\end{equation}
with $\partial_{x}^{2}\rho_x:=\partial^{2}\rho_x/\partial x^{2}$, $\rho_x\rho_{x+\mathrm{d}x}$ can be expressed by 
\begin{equation}
\rho_x\rho_{x+\mathrm{d}x}=\rho^{2}_x+\rho_x\partial_{x}\rho_x\mathrm{d}x
+\frac{1}{2}\rho_x\partial_{x}^{2}\rho_x(\mathrm{d}x)^{2},
\end{equation}
and $\rho^{2}_{x+\mathrm{d}x}$ can be expressed by 
\begin{align}
\rho^{2}_{x+\mathrm{d}x}=&\rho^{2}_x+\rho_x\partial_{x}\rho_x\mathrm{d}x
+\left(\partial_{x}\rho_x\right)\rho_x\mathrm{d}x +\left(\partial_{x}\rho_x\right)^{2}(\mathrm{d}x)^{2} \nonumber \\
&+\frac{1}{2}\rho_x\partial_{x}^{2}\rho_x(\mathrm{d}x)^{2}
+\frac{1}{2}\left(\partial_{x}^{2}\rho_x\right)\rho_x(\mathrm{d}x)^{2}.
\end{align}

Utilizing the expressions above, one can further have 
\begin{align}
&\sqrt{\left[1-\mathrm{Tr}\left(\rho^{2}_x\right)\right]
\left[1-\mathrm{Tr}\left(\rho^{2}_{x+\mathrm{d}x}\right)\right]} \nonumber \\
&=1-\mathrm{Tr}\left(\rho^{2}_x\right)-\mathrm{Tr}\left(\rho_x\partial_{x}\rho_x\right)\mathrm{d}x \nonumber \\
&-\frac{1}{2}\left\{\mathrm{Tr}\left(\rho_x\partial_{x}^{2}\rho_x\right)\!+\!\mathrm{Tr}\left((\partial_{x}\rho_x)^{2}\right)
\!+\!\frac{\left[\mathrm{Tr}\left(\rho_x\partial_{x}\rho_x\right)\right]^{2}}
{1-\mathrm{Tr}\left(\rho^{2}_x\right)}\right\}\mathrm{d}x^{2}.
\end{align}
Hence, $g(\rho_x,\rho_{x+\mathrm{d}x})$ can be calculated as
\begin{equation}
1-\frac{1}{2}\left\{\mathrm{Tr}\left((\partial_{x}\rho_x)^{2}\right)
+\frac{\left[\mathrm{Tr}\left(\rho_x\partial_{x}\rho_x\right)\right]^{2}}
{1-\mathrm{Tr}\left(\rho^{2}_x\right)}\right\}(\mathrm{d}x)^{2}.
\end{equation}
Based on Eq.\,(\ref{eq:app_sub-qfi}), the sub-QFI for a mixed state can then be expressed by 
\begin{equation}
F_{\mathrm{sub}}
=2\mathrm{Tr}\left((\partial_{x}\rho_x)^{2}\right)
+\frac{1}{2}\frac{\left[\partial_{x}\mathrm{Tr}\left(\rho^{2}_x\right)\right]^{2}}
{1-\mathrm{Tr}\left(\rho^{2}_x\right)}.
\end{equation}

\subsection{Equivalence with the QFI for pure states}
\label{sec:apx_subQFI_pure}

When $\rho_x$ is a pure state, the superfidelity reduces to $g(\rho_{1},\rho_{2})=\mathrm{Tr}(\rho_{1}\rho_{2})$, and 
the corresponding sub-QFI can be expressed by 
\begin{equation}
F_{\mathrm{sub}} = 2\Tr((\partial_x \rho_x)^2).
\end{equation}
Utilizing the equation of the SLD (denoted by $L_x$) $\partial_x \rho_x = (\rho_x L_x + L_x \rho_x)/2$, the expression 
above becomes 
\begin{equation}
F_{\mathrm{sub}}= \Tr(\rho_x L_x \rho_x L_x + \rho_x^2 L^2_x).
\end{equation}
For a pure state $\rho_x=\ket{\psi}\bra{\psi}$, the term $\Tr(\rho_x L_x \rho_x L_x)$ reduces to 
\begin{equation}
\Tr(\rho_x L_x \rho_x L_x) = \bra{\psi}L_x\ket{\psi}^2=\Tr^2(\rho_x L_x)=0.
\end{equation}
Here we used the fact that $\Tr(\partial_x \rho_x) = 0 = \Tr(\rho_x L_x)$. Therefore, 
\begin{equation}
F_{\mathrm{sub}} = \Tr(\rho_x L_x^2) = F(\rho_x),
\label{eq:apx_subQFI}
\end{equation}
namely, the sub-QFI for a pure state is equivalent to the corresponding QFI. 

\subsection{Equivalence with the QFI for two-level states}
\label{sec:apx_subQFI_2level}

Next we show the equivalence between the QFI and sub-QFI for any two-level state. It is already known that the 
sub-QFI is equivalent to the QFI for pure states. Therefore, here we only need to discuss the two-level mixed states. 
The density matrix for a two-level mixed state can be written as 
\begin{equation}
\rho_x = \lambda_0\ket{\lambda_0}\bra{\lambda_0} + \lambda_1\ket{\lambda_1}\bra{\lambda_1},
\end{equation}
where $\ket{\lambda_0}$, $\ket{\lambda_1}$ are the eigenvectors of $\rho_x$ and $\lambda_0$, $\lambda_1$ are 
the corresponding eigenvalues and both nonzero. For a two-level system, the QFI can be calculated as\,\cite{Liu2020} 
\begin{equation}
F = \frac{(\partial_x \lambda_0)^2}{\lambda_0(1-\lambda_0)} + 4(1-2\lambda_0)^2|\bra{\lambda_0}\ket{\partial_x \lambda_1}|^2.
\label{eq:apx_QFI2level}
\end{equation}

In the meantime, the sub-QFI can be calculated with Eq.\,(\ref{eq:apx_subQFI}). In this case, it can be easy to find that 
\begin{equation}
\Tr\left((\partial_x \rho)^2\right) =2(\partial_x \lambda_0)^2+2(1-2\lambda_0)^2|\bra{\lambda_0}\ket{\partial_x \lambda_1}|^2, 
\end{equation}
and 
\begin{equation}
\frac{1}{2}\frac{(\partial_x \Tr(\rho^2))^2}{1-\Tr(\rho^2)} =\frac{(1-2\lambda_0)^2(\partial_x \lambda_0)^2}{\lambda_0(1-\lambda_0)}.
\end{equation}
Thus, the sub-QFI can be expressed by 
\begin{equation}
F_{\mathrm{sub}} = \frac{(\partial_x \lambda_0)^2}{\lambda_0(1-\lambda_0)}+4(1-2\lambda_0)^2|\bra{\lambda_0}\ket{\partial_x \lambda_1}|^2,
\end{equation}
which is equivalent to Eq.\,(\ref{eq:apx_QFI2level}). Hence, the sub-QFI is always equivalent to the QFI for any two-level state. 

\section{Rank of the SLD operator corresponding to a pure state}
\label{sec:qpx_rank}

A pure state $\ket{\psi}$ can be expressed in an orthonormal basis $\{\ket{c_i}\}$ as
\begin{equation}
\ket{\psi} = \sum_{i=1}^{N} c_i \ket{c_i}, 
\end{equation}
where $c_j$ are the coefficients of the basis vectors. In the basis $\{\ket{c_i}\}$, the SLD of a pure state can be expressed by 
\begin{align}
L &= 2\left(\ket{\partial_x \psi}\bra{\psi}+\ket{\psi}\bra{\partial_x \psi}\right) \nonumber \\
&= \sum_{i,j=1}^N 2(\partial_x c_i) c_j^{*} \ket{c_i}\bra{c_j}+\sum_{i,j=1}^N 2 c_i (\partial_x c_j^{*}) \ket{c_i}\bra{c_j}.
\end{align}
Define two vectors $\vb{u}=(\partial_x c_1, \partial_x c_2, \cdots, \partial_x c_N)^{\mathrm{T}}$ and 
$\vb{v}=(c_1, c_2, \cdots, c_N)^{\mathrm{T}}$ in the basis $\{\ket{c_i}\}$, then we have
\begin{equation}
L = 2\vb{u}\vb{v}^\dagger + 2\vb{v}\vb{u}^\dagger.
\end{equation}
Note that $\rank(\vb{u}\vb{v}^\dagger) = \rank(\vb{v}\vb{u}^\dagger) = 1$ where $\rank(\cdot)$ denotes the rank, so $\rank(L)\leq 2$. 
In the meantime, the SLD operator must have an even number of nonzero eigenvalues due to the fact that $\bra{\psi}L\ket{\psi}=0$. Hence, 
its rank has also to be even, which immediately indicates that $\rank(L)=2$.

\end{document}